# Transient nanoscopic phase separation in biological lipid membranes resolved by planar plasmonic antennas

Pamina M. Winkler[†], Raju Regmi[‡,†], Valentin Flauraud[§], Jürgen Brugger[§], Hervé Rigneault[‡], Jérôme Wenger[‡], María F. García-Parajo[*†⊥]

[†] ICFO-Institut de Ciencies Fotoniques, The Barcelona Institute of Science and Technology, Barcelona, Spain;

[‡] Aix Marseille Univ, CNRS, Centrale Marseille, Institut Fresnel, Marseille, France;

[§] Microsystems Laboratory, Institute of Microengineering, Ecole Polytechnique Fédérale de Lausanne, 1015 Lausanne, Switzerland;

[⊥] ICREA, Pg. Lluís Companys 23, 08010 Barcelona, Spain.

**Abstract:**

Nanoscale membrane assemblies of sphingolipids, cholesterol and certain proteins, also known as lipid rafts, play a crucial role in facilitating a broad range of important cell functions. Whereas on living cell membranes lipid rafts have been postulated to have nanoscopic dimensions and being highly transient, the existence of a similar type of dynamic nanodomains in multicomponent lipid bilayers has been questioned. Here, we perform fluorescence correlation spectroscopy on planar plasmonic antenna arrays with different nanogap sizes to assess the dynamic nanoscale organization of mimetic biological membranes. Our approach takes advantage of the highly enhanced and confined excitation light provided by the nano-antennas together with their outstanding planarity to investigate membrane regions as small as 10 nm in size with microsecond time resolution. Our diffusion data are consistent with the coexistence of transient nanoscopic domains in both the liquid-ordered and the liquid-disordered microscopic phases of multicomponent lipid bilayers. These nanodomains have characteristic residence times between 30 μs and 150 μs and sizes around 10 nm, as inferred from the diffusion data. Thus, although microscale phase separation occurs on mimetic membranes, nanoscopic domains also coexist, suggesting that these transient assemblies might be similar to those occurring in living cells which in the absence of raft-stabilizing proteins are poised to be short-lived. Importantly, our work underscores the high potential of photonic nano-antennas to interrogate the nanoscale heterogeneity of native biological membranes with ultra-high spatiotemporal resolution.

The spatiotemporal lateral organization and the biological function of the eukaryotic plasma membrane are intricately interlaced at the nanoscale. It is well accepted that the landscape of the cell membrane is highly heterogeneous and shaped by a variety of lipids and proteins that differ in their physico-chemical properties. In the plane of the membrane, lateral heterogeneities resulting from the formation of specialized regions enriched in sphingolipids, cholesterol and specific proteins are commonly known as lipid rafts.[1–4] These lipid assemblies are thought to constitute a tightly packed, short-range, liquid-ordered (Lo) phase coexisting with a more liquid-disordered (Ld) phase within the surrounding fluid membrane.[5–7] While the existence of phase separation in the plasma membrane has been debated for many years, a large number of recent experimental data is convincingly demonstrating that lipid rafts in living cell membranes have nanoscopic dimensions and are highly dynamic.[8–13] Importantly, lipid rafts play a crucial role in many cellular processes that include signal transduction, protein and lipid sorting and immune response amongst others.[2,5,10,14,15] Understanding the formation mechanism and properties (e.g. size, composition) of lipid rafts and relating their structure to their functional role is of paramount interest.

Model lipid membranes represent a simple mimetic system that recapitulates some of the most important features of biological membranes, i.e., spatiotemporal compartmentalization and lipid phase separation. On the microscopic scale, ternary lipid membranes composed of unsaturated phospholipids, saturated sphingolipids and cholesterol separate into coexisting Ld and Lo phases, which can be resolved by diffraction-limited optics.[16–18] The large microscopic size and the stable nature of Lo domains observed on mimetic membranes strongly contrasts with the highly transient and nanoscopic size of lipid rafts inferred on living cells. Interestingly, while there is a continuous push for resolving nanoscopic lipid domains in the plasma membranes of living cells, recent works suggest that the microscopically homogeneous Lo and Ld phases on lipid model membranes might in fact also be heterogeneously organized at the nanometer scale.[6,19,20]

The possibility that nanoscale lipid heterogeneities also exist within the supposedly homogenous Lo and Ld phases of artificial membranes is intriguing and of particular interest as they might be the underlying basis for lipid raft formation in living cells. Indeed, earlier work from Hancock and co-workers predicted that in the absence of stabilizing proteins, the size of lipid nanoassemblies would be smaller than 10 nm in diameter and short-lived, with lifetimes below 1 ms.[6,21] Consistent with this hypothesis, recent deuterium-based nuclear magnetic resonance (d-NMR) experiments revealed the presence of cholesterol and sphingolipids in the Ld phase as well as of unsaturated lipids in the Lo phase, suggesting that nano-sized clusters may exist in both phases.[19] Fluorescence correlation spectroscopy (FCS) in combination with stimulated emission depletion (STED) nanoscopy has been recently applied to study the nanoscale dynamics occurring in Lo and Ld phases of ternary lipid-cholesterol mixtures with a spatial resolution of 40 nm. The results showed fully homogenous Lo and Ld phases with no evidence on the occurrence of nanoscopic domains at the tested spatial scales (40-250 nm).[22] However, it is possible that nanoscale assemblies smaller than the 40 nm STED resolution could not be detected. In contrast to these results, recent high-speed single particle tracking (SPT) of 20 nm gold beads attached to individual lipids showed anomalous diffusion on the Lo phase consistent with the occurrence of nanoscale heterogeneities,  while homogeneous lipid diffusion was observed on the Ld phase.[20] The estimated sizes of the nanodomains on the Lo phase varied between 20 to 40 nm with lipid trapping times inside the domains below 1 ms.

A different, and potentially powerful approach to investigate dynamic nanoscale heterogeneities of lipid bilayers is provided by the use of plasmonic antennas. These metal nanostructures enhance and confine light down to nanoscale dimensions, sustaining localized hotspot regions of the excitation light.[23–26] Moreover, when combined with FCS, single molecule detection at ultra-high sample concentrations with microsecond time resolution can be obtained, both in solution and living cell membranes.[27–31] However, in most antenna desings the region of maximum field localization and enhancement (i.e., hotspot) is buried into the nanostructure, and thus difficult to access. Recently, we overcame this drawback by fabricating *in-plane* dimer antenna arrays where the gap region is located at the sample surface.[32] This design provides direct accessibility to the antenna hotspot region and drastically improves the optical performance to yield fluorescence enhancement factors of up to $10^4-10^5$ together with nanoscale detection volumes in the zeptoliter range.[32] Here, we take advantage of the strong optical confinement occurring on plasmonic antenna arrays together with their remarkable planarity to inquire on the nanoscale dynamics of multicomponent lipid bilayers. Our results reveal for the first time the coexistence of transient nanoscopic domains in both the Lo and Ld phase, in the microsecond scale and with characteristic sizes below 10 nm. These nanoscale assemblies might be reminiscent to those naturally occurring in living cells, which in the absence of raft-stabilizing proteins, are expected to be highly transient.

**Results and discussion**

Model lipid bilayers with different compositions were prepared on glass coverslips or on top of antenna substrates following a modified protocol from Ref. 33 and explained in the Methods section. Bilayers were composed of the unsaturated phospholipid 1,2-dioleoyl-sn-glycerol-3-phosphocholine (DOPC) alone, DOPC in combination with sphingomyelin (18:0 SM) (1:1 molar proportions) and of ternary mixtures of DOPC, SM (1:1) with addition of 10 or 20 mol% cholesterol (Chol). The different bilayers were labeled with the lipophilic fluorescent dye DiD which preferentially partitions in the Ld phase.[34,35] The quality of the glass-supported bilayers was assessed by FCS measurements using a diffraction-limited confocal microscope. The obtained diffusion values are in good quantitative agreement with values reported for similar lipid mixtures[34] (see Supporting information Figure S1, Table S1 and related discussion).

To investigate the existence of nanoscale heterogeneities in the different lipid mixtures, we used the same preparation protocol to create bilayers on top of the *in-plane* antenna substrates as schematically illustrated in Figure 1a. The antenna design consisted of gold dimers of 80 nm in diameter separated by nanogaps of different sizes (from 10 nm to 45 nm) and surrounded by nano-apertures to further constrain the excitation area and reduce background contribution from fluorescent molecules diffusing outside the antenna hotspots (Supporting information Figure S2a).[32] For the experiments reported here we took advantage of the *in-plane* antenna geometry to have access to the maximum spatial confinement at the gap regions. In addition, considering that the height difference between the gold dimers and the filling polymer at the hotspot measurement site is below 1nm (Supporting information Figure S2b), we regard these substrates as of excellent planarity and thus suitable for membrane studies.

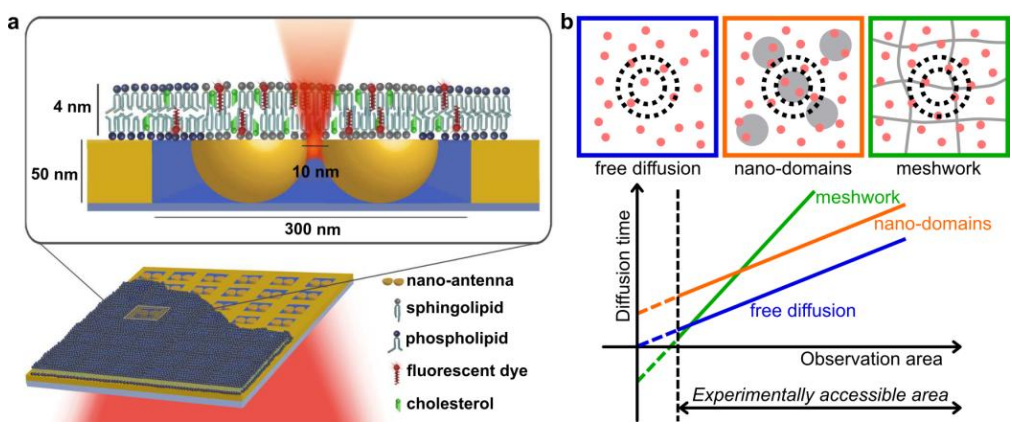

**Figure 1. Biological lipid membranes probed by *in-plane* plasmonic antenna arrays.** (a) Schematics of the experimental measurements. Bilayers of different lipid compositions are deposited on top of *in-plane* antenna arrays. Each antenna consists of a gold dimer separated by a nanogap and embedded in a nano-aperture. Individual antennas are excited using a confocal set-up. The confined and enhanced field at the antenna hotspot excites individual DiD molecules embedded in the bilayer. Fluorescence fluctuations arising from the passage of molecules through the hotpot of the antenna are recorded and autocorrelated in time to generate ACF curves. (b) FCS diffusion laws to extract the type of diffusion exhibited by molecules in biological membranes. The excitation area (dashed circles in the upper sketch) is varied and the diffusion time of molecules (red dots) is plotted as a function of the excitation area (lower plot). The vertical dashed line in the plot denotes the experimentally accessible area. Three different plots can be obtained reflecting the type of diffusion experienced by the molecule. Continuous lines represent measured values on the experimentally accessible illumination areas. Dashed lines correspond to the extrapolation of the curves through the $y$-intercept. The plots are fitted to $\tau(w^2) = t_0 + w^2/(4 \times D)$, where $w^2$ corresponds to the illumination area, D to the effective diffusion coefficient and $t_0$ corresponds to the $y$-intercept. Free diffusion is characterized by $t_0 = 0$; $t_0 > 0$ denotes constrained diffusion due to nanodomains and $t_0 < 0$ corresponds to constrained diffusion due to meshwork obstacles.

FCS is a single-molecule sensitive technique that records intensity fluctuations of fluorophores as they transit through the illumination volume. The obtained intensity fluctuations are temporally correlated and the resulting autocorrelation function (ACF) provides quantitative information on the diffusion behavior and absolute concentration of the detected molecules. To assess the presence of nanoscale heterogeneities in the lipid membranes that would alter the diffusion of the dye, we carried out FCS experiments on antennas of different gap sizes to record the evolution of the diffusion time as a function of the illumination area. This approach exploits the so-called FCS diffusion laws as introduced in Ref. 36. A diffusion law or diffusion plot is a curve representing the progression of the molecular diffusion time for increasing detection areas, which generally follows a straight line (Figure 1b). The slope of the curve represents the effective diffusion coefficient describing the long-range mobility of the molecule, while the $y$-intercept $t_0$ on the time axis indicates whether the molecule undergoes free Brownian diffusion ($t_0 = 0$), or is dynamically partitioning into nano-domains ($t_0 > 0$) or trapped in a molecular meshwork ($t_0 < 0$).[36,37]

In this work, we used nano-antennas of nominally 10, 30 and 45 nm gap sizes. The sizes of the illumination areas were estimated based on the gap sizes as measured on TEM images (Supporting information Figure S3), and numerical simulations (Supporting information Figure S2c-e and Methods). In addition, we performed calibration measurements of the area sizes using the Alexa Fluor 647 dye in solution for five different antenna gaps (Supporting information Figure S4 and Methods). From the calibration curves, we experimentally determined values of (300 ±50) nm$^2$, (1080 ±80) nm$^2$ and (2025 ±110) nm$^2$ for the 10, 30 and 45nm gap antennas respectively, which are in excellent agreement with the calculated values. The sizes of the illumination areas are between one to two orders of magnitude smaller than the ones of confocal excitation, underscoring the extreme light confinement afforded by plasmonic antennas.

Samples were excited by focusing the incoming laser light ($\lambda$=640nm, ~2 kW/cm$^2$) onto individual antennas using a water-immersion objective (NA=1.2). Under these excitation conditions, the temperature increase at the antenna hotspots due to optical heating was estimated to be only within 1-3 K.[38] The fluorescence signal was collected in reflection mode by the same objective, filtered from the excitation light and sent to two single photoncounting APD detectors. As the antennas show a polarization-dependent response, we used excitation polarization parallel to the antenna gaps to achieve maximum field enhancement and confinement.[32] Fluorescence fluctuations arising from the diffusion of DiD in the bilayers were recorded for at least 30 seconds at each individual antenna and the resulting normalized ACFs were calculated.

Representative fluorescent intensity time traces of DiD diffusion in a pure DOPC membrane over three different antenna gap areas are shown in Figure 2a, together with enlarged views of representative single bursts. The burst duration increases with gap area, confirming that the detected signal arises from the excitation of the dye at the gap regions. This is further substantiated by the normalized ACFs obtained for different gap sizes and compared to confocal measurements (Figure 2b). To extract the diffusion times from individual ACF curves, we performed two-component 2D Brownian fittings to account for both, direct excitation from the gap region ($\geq$ 55-90% of the weighted amplitude) and residual excitation of DiD diffusing through the nano-aperture (see Eq. (1) in the Methods section). Results of the main component of the fitted curves shown in Figure 2b render $\tau_{DOPC}$ values of (6±1) $\mu$s, (25±3) $\mu$s and (71±25) $\mu$s for the respective gap areas of (300, 1080, and 2025) nm$^2$ compared to $\tau_{DOPC}$ = 3.5 ms for confocal excitation. The complete results of the fittings and relative contributions of the gap and nano-aperture excitation are shown in the Supporting information Table S2.

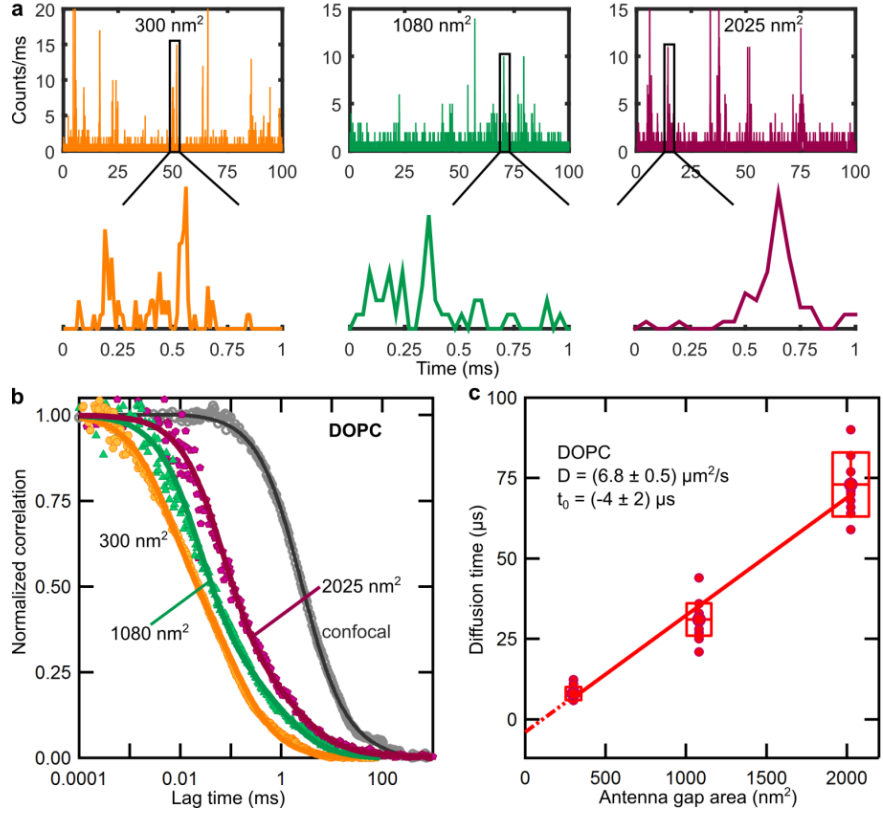

**Figure 2. FCS measurements in pure DOPC bilayers using *in-plane* plasmonic antenna arrays of different gap sizes.** (a) Representative fluorescence intensity time traces of DiD embedded in a pure DOPC bilayer for three different antenna gap areas together with enlarged views of representative bursts. (b) Normalized ACF curves as obtained from different antenna gap areas and by confocal excitation. (c) Diffusion times as extracted from the ACF fitting as a function of the antenna gap area. Each dot corresponds to an individual ACF measurement in a single antenna. Number of measurements: 13, 8 and 12 for antenna gap areas of 300 nm$^2$, 1080 nm$^2$ and 2025 nm$^2$ respectively on five different samples. Fitting by orthogonal distance regression (red line) has been performed through the mean diffusion time values of each respective antenna gap area minimizing the error for the gap area and ± 1std in diffusion time (horizontal and vertical line of the red box, respectively).

Diffusion times obtained from multiple measurements on individual antennas as a function of the probed gap are shown in Figure 2c. The mean values of the diffusion times versus gap area could be well-fitted to a straight line with an intercept close to the zero-origin point, indicating that $\tau_{\text{DOPC}}$ scales linearly with the gap area, consistent with free Brownian diffusion of the dye in the pure DOPC membrane. The slope of the fitting rendered a diffusion coefficient of (6.8±0.5) $\mu m^2$/s which compares well to our confocal measurements (see Supporting information Table S1) and values reported elsewhere.[34]

To further validate that these short diffusion times arise from the strong optical confinement occurring at the surface of the gap regions, we performed experiments on similar antennas with excitation light perpendicular to the antenna gap. In these conditions, the antennas are not resonantly excited and the excitation field essentially corresponds to that of the surrounding nano-apertures alone. Accordingly, the fluorescence signal is much weaker and the ACF curves look much noisier (Supporting information Figure S5). ACF curves for perpendicular excitation could be fitted with a single Brownian diffusion component yielding much longer transient times (1.2 ms – 1.9 ms), which are close to the values obtained upon excitation of the nano-aperture alone (i.e., with no antennas inside). These values are also comparable to the $t_2$ values of the second contribution obtained from the two-component fitting performed on the ACF curves for parallel polarization excitation of the antennas (see also Supporting Information Table S2).

Overall, these results on pure DOPC bilayers validate the application of plasmonic antennas to record the diffusion of individual molecules in lipid bilayers with microsecond time resolution and clearly demonstrate their nanoscale excitation confinement. Additionally, our results show that DOPC bilayers are homogenous down to the nanoscale. It further confirms that the antenna substrates supporting the bilayers are of extreme flatness and quality as no hindering effects on the dye diffusion were observed, neither on the DOPC bilayers nor in solution (see also Figure 5a,b).[32] It has been recently reported that molecular pinning and interleaflet membrane coupling effects leading to deviations from free Brownian diffusion at the nanoscale are influenced by the properties of the substrate, e.g. plasma-cleaned glass vs. mica support.[39] We believe that these effects are not present in our DOPC measurements since: a) we observed Brownian diffusion of DOPC down to the nanoscale, with a $t_0$ intercept close to zero; b) the gold antenna substrates have been pre-treated with UV/ozone plasma cleaning immediately prior to the bilayer deposition (see Methods). This treatment leads to a chemically inert and hydrophilic gold surface.[40]

To shed light on the diffusive behavior of lipid membranes composed of binary and ternary mixtures, lipid membranes composed of DOPC:SM (1:1) alone and with the addition of 10 and 20 mol% of Chol were prepared on top of the antenna array substrates and probed by means of FCS. Figure 3a shows characteristic fluorescent intensity time traces of DiD diffusing across the smallest gap antenna (300 $nm^2$ hotspot area) in DOPC:SM bilayers and on a ternary mixture containing 20 mol% of Chol. In the presence of Chol, stable macroscopic phase separation occurs, so that depending on the antenna location with respect to the membrane, different fluorescence trajectories are recorded, either probing the Ld phase (Figure 3a, middle trajectory) or the Lo phase (Figure 3a, right trajectory). The enlarged views of single bursts of the individual trajectories show increasing burst durations for the binary and ternary mixtures as compared to DOPC (Figure 2a) consistent with slower diffusion of the dye in these lipid mixtures. ACF curves for DOPC:SM and the ternary mixture of DOPC:SM with 20 mol% Chol for both phases (Ld and Lo) are depicted in Figure 3b. As for the DOPC measurements, we fitted all the ACF curves using a two-component 2D Brownian diffusion model. We also attempted to fit the two ACF curves of the ternary mixture to an anomalous diffusion model. However, considering that the anomaly parameter $\alpha$, did not improve the fitting and rendered $\alpha$ values larger than 1, we opted for the use of the two-component 2D Brownian diffusion model. The derived diffusion times at the gap regions for the different lipid compositions resulted in $\tau_{DOPC:SM} = (36\pm4)$ µs, and $\tau_{Ld} = (39\pm9)$ µs and $\tau_{Lo} = (229\pm30)$ µs for the 20 mol% Chol ternary lipid mixture. The diffusion times obtained from multiple ACF curves on different 300 $nm^2$ gap areas for all the different lipid mixtures are summarized in Figure 3c. While DiD in DOPC shows the shortest diffusion time, addition of SM slowed down the dye diffusion, consistent with the confocal results. The longest diffusion times were observed for the Lo phase of the ternary mixtures. Interestingly, the diffusion times of DiD in the Ld phase are significantly longer than those obtained in the pure DOPC bilayer, which may already indicate the presence of transient nanoassemblies of Chol in the Ld phase.

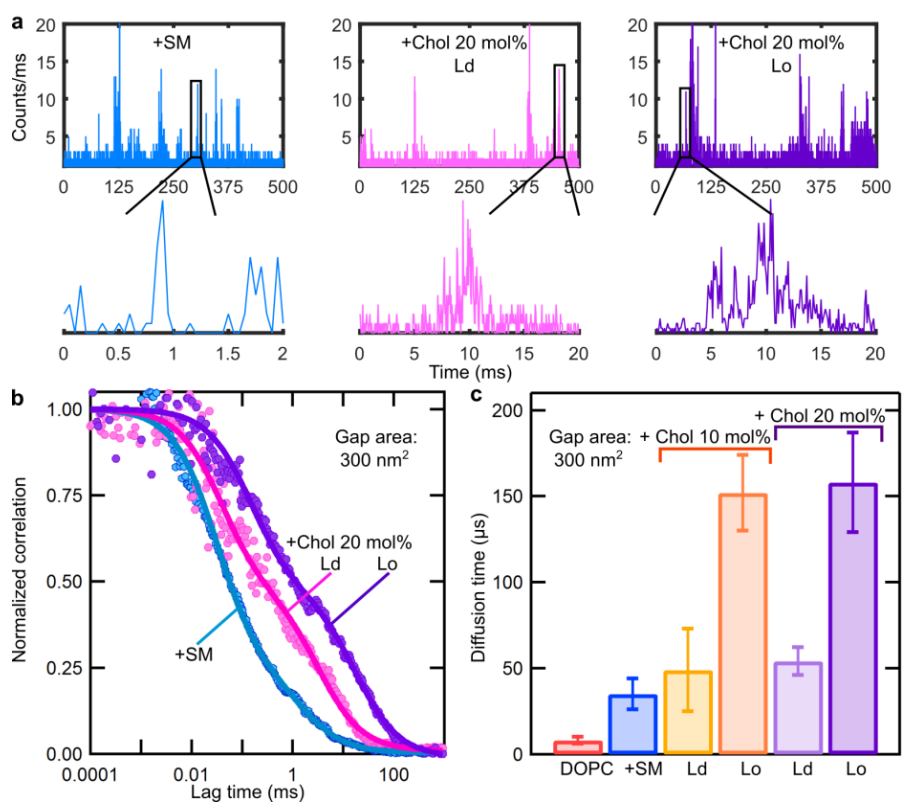

**Figure 3. FCS measurements recorded in nanogaps of 300 nm² area for binary and ternary lipid mixtures. (**a) Fluorescence intensity time traces of DiD diffusion in DOPC:SM (1:1) (left, blue) and DOPC:SM (1:1) + Chol (20 mol%) in the Ld (middle, magenta) and Lo (right, purple) phases. Enlarged views of single bursts are also depicted for visual comparison. (b) Normalized ACF curves for the different lipid mixtures. (c) Mean diffusion times of the four probed lipid membranes ±1std obtained for the smallest gap area (300 ± 48) nm². Number of measurements: 20 for DOPC; 19 for DOPC:SM; 18 and 15 for Ld and Lo, respectively with Chol 10 mol%; and 17 and 15 for Ld and Lo, respectively with Chol 20 mol%. Between 5 and 10 different antennas of 300 nm² area on 4 to 5 different samples were used for each lipid composition.

To gain more insight into these results we measured the diffusion times for all the lipid mixtures for different antenna gap sizes and over multiple antennas. The data were fitted through the mean diffusion time values to obtain diffusion laws for each lipid composition (). Two main parameters can be directly extracted from the fitting, i.e., the effective diffusion coefficient which is calculated from the slope of the curves and the *y*-intercept of the fitting at zero gap areas ($t_0$). In the case of DOPC and DOPC:SM bilayers, the intercepts of the fitting cross the origin at nearly zero diffusion times (Figure 4a) consistent with free diffusion of the dye in these lipid bilayers, albeit the diffusion in the DOPC:SM mixture is significantly slowed down as compared to the pure DOPC membrane due to the tighter packing of saturated SM. In marked contrast, positive $t_0$ values are obtained for both the Ld and Lo phases in the ternary mixtures containing 10 and 20 mol% Chol indicating that the diffusion of the dye is not Brownian (Figure 4b,c). Note that the Ld and the Lo phases for the two ternary lipid model membrane mixtures containing 10 and 20 mol% of Chol as shown in Figure 4b,c can clearly be distinguished based on the markedly different diffusion times (see Supporting information Figure S6).

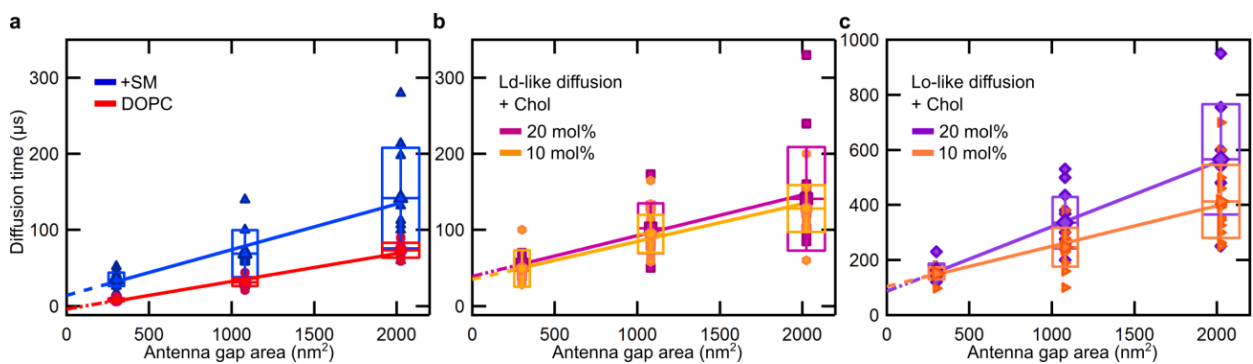

**Figure 4. FCS diffusion laws extended to the nanoscale for different lipid mixtures.** Diffusion times versus antenna gap areas in (a) DOPC and DOPC:SM bilayers. (b) Ld phase for 10 and 20 mol% Chol; and (c) Lo phase for 10 and 20 mol% Chol. The colored dots represent diffusion times acquired from FCS on individual nano-antennas of different gap sizes, while the solid lines are fits through the mean values. Measurements were performed on at least 15 different nano-antennas for each lipid composition on five different samples.

Results of the effective diffusion coefficients together with the $t_0$ values for all lipid compositions are shown in Figure 5a,b. For the ternary mixtures containing cholesterol, the diffusion coefficients in the Ld phase are significantly slower than the one of pure DOPC (mean values of 5.1 and 4.7 μm²/s for 10 and 20 mol% Chol respectively, compared to 6.8 μm²/s for DOPC) (Figure 5a), which is consistent with the longer diffusion times reported in Figure 3c. Moreover, the positive $t_0$ values indicate a deviation from Brownian diffusion due to the presence of heterogeneities in the Ld phase caused by the occurrence of nanodomains (Figure 5b). These nanoscopic domains are most probably formed by the presence of SM and Chol in the Ld phase, which reduce the effective diffusion of the dye as compared to the pure DOPC bilayer. A similar and even more pronounced trend is also observed in the Lo phase, where the dye experiences very slow effective diffusion and strong deviation from Brownian motion, indicating also the presence of heterogeneities and nanodomains in this phase (Figure 5a,b). Altogether, these results clearly indicate the occurrence of transient nanoscale heterogeneities in both phases of lipid model membranes which have remained so far beyond the detection limits of conventional microscopy.

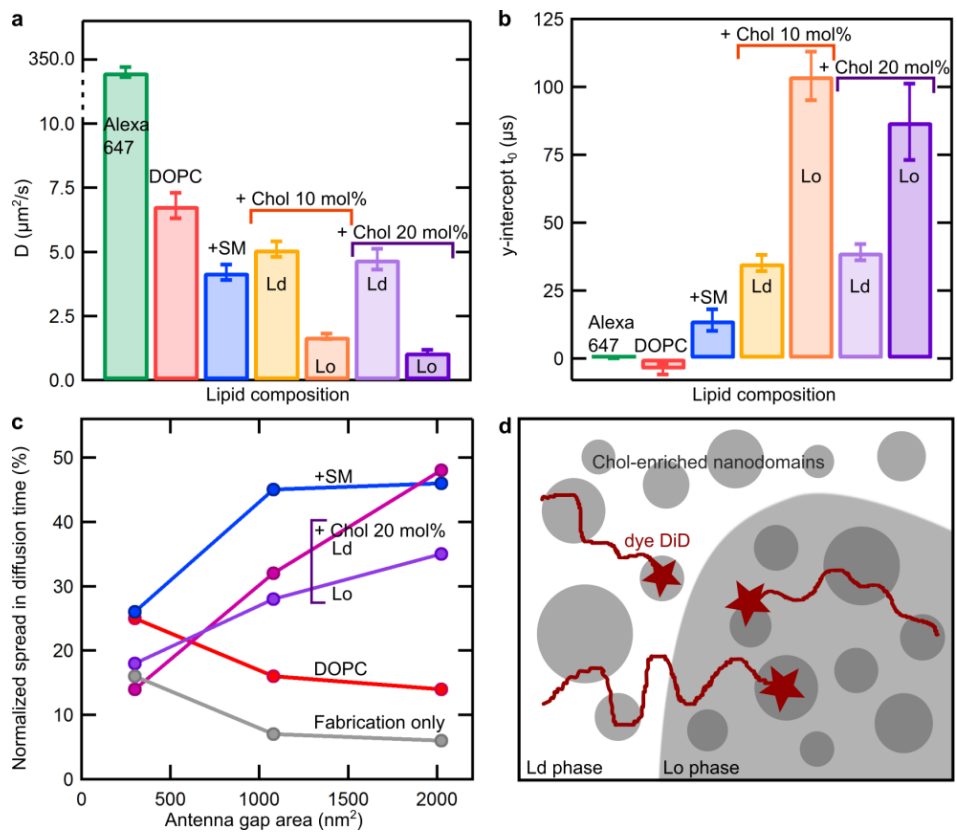

**Figure 5. In-plane plasmonic nanogap antennas reveal nanoscopic heterogeneities in the Ld and Lo phases of biological lipid membranes.** (a) Effective diffusion coefficients $D$ and (b) $y$-intercepts $t_0$ (with ±1std for the errors) as extracted from FCS diffusion laws shown in Figure 4 for the different lipid membrane compositions. Data of the dye Alexa Fluor 647 is included for comparison and showing its expected Brownian free diffusion in solution, i.e., $t_0 = 0$ (taken from Ref. 32). (c) Normalized spread in diffusion times (± 1std/mean × 100%) as a function of the antenna gap area for three different lipid compositions. The gray dots and line correspond to the spread in diffusion times resulting from fabrication inaccuracies of the antenna gaps (variation in

gap size/mean gap size, see also Figure S2) (d) Sketch illustrating the presence of dynamic nanoscopic domains inside microscopic Lo and Ld phases of lipid mixtures containing cholesterol.

Further evidence on the heterogeneity within the Lo and Ld phases for ternary lipid bilayers containing Chol can be inferred from the spread of the diffusion times for similar antennas as a function of the gap area, which we quantified as the deviation from the mean diffusion time. Since the metal thin-film morphology affects the fabrication process of the antennas and leads to variations on the real gap areas, we first estimated the spread in diffusion times arising solely from the differences in gap areas (see also Supporting information, Figure S2). As expected, the spread in diffusion times reduces with increasing gap area as the fabrication process becomes more accurate and smaller variations in the gap sizes are obtained (Figure 5c). In pure DOPC bilayers, DiD shows a spread of the diffusion times that reduces with increasing gap size and essentially reflects the effect of fabrication inaccuracies (Figure 4a and Figure 5c). In striking contrast, the spread of the diffusion times for the binary and ternary compositions increases significantly with gap size (Figure 4a-c and Figure 5c) suggesting that the dye samples heterogeneous regions of faster and slower diffusion. These results directly correlate with the positive $t_0$ values measured, indicating the occurrence of heterogeneities in both the Lo and Ld phases due to nanoscopic domains that diffuse through the hotspot gap area during the measurements. Notably, in the case of DOPC:SM, large spreads of the diffusion times are observed for both the (1080 and 2025) nm$^2$ antennas, although diffusion is largely free and homogeneous (notice that $t_0$ for the DOPC:SM mixture is slightly positive, i.e., (14±4) μs). We interpret these results as the occurrence of extremely transient local concentrations of SM within the otherwise homogenous DOPC layer that lead to differences in transient diffusion times but without the occurrence of detectable nanodomains.

The characteristic residence time $\tau_{res}$ of the dye in a single nanodomain can be estimated from $t_0$ through the relation $t_0 \approx 2\alpha\tau_{res}$, since the confinement time is much larger that the free diffusion time inside the domain,[36] and where $\alpha$ corresponds to the partition coefficient of the dye in the Ld and Lo phases.[36] In the case of DiD, the values of $\alpha$ correspond to 0.66 for the Ld phase and 0.33 for the Lo phase evaluated for microscopically phase-separated domains[34] and we assume no significant deviations at the nanoscale. Therefore, the residence times inside the different nanoscopic domains yield values for the Ld phase of $\tau_{res}$= (27 ± 3) μs and (30 ± 3) μs for 10 and 20 mol% Chol respectively, and for the Lo phase of (158 ± 9) μs and (132 ± 14) μs for 10 and 20 mol% Chol, respectively. Although from our measurements we cannot directly estimate the sizes of these nanoscopic domains, the large spread of the diffusion times observed for the two larger gap areas (1080 nm$^2$ and 2025 nm$^2$) indicates that these gap areas already probe different nanoscopic regions during our measurements. On the other hand, the spread on the diffusion times reduces for the smallest antennas (300 nm$^2$) and are similar to the variations obtained from the fabrication procedure (Figure 5c). Reduced variations in the diffusion times implies that the sizes of the nanoscopic domains become comparable to the illumination area, which for the smallest antennas is 10 nm in size.

Overall, our results provide compelling evidence for the existence of nanoscopic domains in both the Ld and Lo phases of multicomponent lipid bilayers, with sizes around 10 nm and short transient residence times around 30 μs for the Ld, and 150 μs for the Lo phase (Figure 5d). Nanoscopic domains in the Lo phase have also been recently detected by high-speed single particle tracking reporting on sizes and residence times that agree remarkably well to the values reported in this work using a completely different technique and experimental approach.[20] Interestingly, earlier stochastic models predicted that lipid rafts in living cells would have to be small (≤ 14nm) with an average residence time of ∼ 60 μs in order to facilitate intermolecular collisions between different proteins.[21] These values come very close to our experimentally measured values. Moreover, molecular dynamics simulations of lipid diffusion within rafts and non-raft domains have predicted the existence of transient clusters with sizes around 10 nm and lifetimes in the microsecond range,[41] once more in excellent agreement with our experimental values. Thus, although the plasma membrane of living cells has a much higher complexity than our model lipid bilayers, we propose that the nanoscopic and highly transient domains detected in this model system might exhibit similar biophysical properties as those predicted in living cells.

Strikingly, our results show also the occurrence of nanoscopic domains in the Ld phase, which to our knowledge, have not been detected before. However, there is ample experimental data that supports their existence. Indeed, earlier FRET measurements showed heterogeneities in the Ld phase that depended on the amount of cholesterol and persisted even at physiological temperatures, hinting towards the existence of nanoscale domains in the Ld phase.[42,43] Moreover, recent d-NMR experiments showed a relatively large percentage of saturated lipids (17%) and Chol (20%) in the Ld phase, which is supposedly composed by only DOPC.[19] The presence of SM and Chol will lead to nanoscopic phase separation within the Ld phase with sizes and lifetimes that would most probably depend on the amount of Chol and SM.[42,43] In our experiments, we find that these nanoscopic domains are extremely short-lived with residence times below 30 μs, being probably the reason why they have not been detected before. In fact, the temporal resolution of the high-speed SPT experiments reporting the presence of nanoscopic domains on the Lo phase was 20 μs, which is not enough to detect transient confinements around 30 μs.[20] By combining nanoscale observation areas as provided by plasmonic antennas with microsecond time resolution as afforded by FCS we have been able to resolve transient nanoscopic domains coexisting in both Ld and Lo phases of mimetic lipid membranes. It is worth mentioning that recent advancements in interferometric scattering (iSCAT) microscopy allow nowadays nanometer localization precision together with microsecond time resolution by the use of 20-40nm gold beads as labeling probes.[20,39,44] Provided that careful controls on potential labeling artifacts and background characterization and removal are performed, iSCAT also constitutes an attractive tool to investigate dynamic biophysical processes at the nanometer scale.

In summary, we have exploited *in-plane* plasmonic nanogap antenna arrays to investigate the lateral organization of lipid model membranes at the nanoscale with microsecond time resolution. The suitability of *in-plane* antenna arrays has been validated on pure DOPC bilayers obtaining free diffusion over the length scales investigated (down to 10 nm), consistent with a homogenous lipid distribution. Free diffusion was also observed on DOPC:SM binary mixtures, although a large spread of the diffusion times was retrieved indicating local fluctuations of SM within larger areas of solely DOPC, but without formation of detectable domains that would constrain dye diffusion. Addition of cholesterol resulted in microscopic phase separation and the formation of transient nanoscopic domains in both the Lo and Ld phases, with sizes below 10 nm and lifetimes in the microsecond time scale. Since the basic biochemistry operating in lipid model membranes is similar to the one in the plasma membrane, we propose that the nanoscopic domains detected here might correspond to the unstable lipid rafts predicted to exist in living cell membranes. The ultra-high spatiotemporal resolution provided by *in-plane* plasmonic nanogap antennas holds great promise to unequivocally reveal their existence in living cells.

**Methods**

*Lipids*

1,2-dioleoyl-*sn*-glycero-3-phosphocholine (DOPC) and N-stearoyl-D-*erythro*-Spingosylphosphorylcholine 18:0 (SM) were purchased from Avanti (Avanti Polar Lipids, Inc.). Cholesterol (Chol) ≥99% was purchased from Sigma-Aldrich and the fluorescent dye $DiIC_{18}(5)$ solid (DiD) from Molecular Probes, Life Technologies Corporation.

*Fabrication of in-plane plasmonic antenna arrays.*

In-plane dimer antenna arrays with gaps of different sizes were fabricated in gold onto glass-coverslips following a procedure described in Ref.32 In brief, the antenna fabrication process was performed on a thin silicon nitride layer deposited on silicon substrates that provided suitable electrical conductivity and chemical stability for the subsequent process steps. Hydrogen-silsesquioxane (HSQ) resist was first spun and then patterned using electron-beam lithography on top of the substrate. A thin layer of gold (50 nm thick) was then deposited by electron beam evaporation over the patterned HSQ structures followed by a planarization step by flowable oxide spin coating. After etching back by Ar-based ion beam etching and removal of the remaining HSQ, the final antenna dimers were stripped from the substrates using a UV curable polymer. This step provided flipped over antennas with accessible gap regions onto optically transparent microscope coverslips. Prior to lipid bilayer deposition, the antenna substrates were carefully cleaned with ethanol, MilliQ water rinsing and UV light exposure for 1 minute followed by 3 minutes of ozone treatment.

*Preparation of lipid model membranes and substrate support*

For the preparation of glass-supported lipid bilayers, glass coverslips were extensively cleaned with acetone, ethanol, ultrasonic bath and in a UV/ozone cleaner with intermediate rinsing steps with pure MilliQ water and immediately used afterwards. Lipid mixtures of DOPC, DOPC:SM (1:1) and DOPC:SM:Chol (10/20 mol%) dissolved at 1 mg/mL in chloroform:methanol (9:1) together with 0.01 mol% of the fluorescent dye DiD were mixed in small glass bottles on ice at 4ºC and immediately deposited on the cleaned coverslips or antenna substrates. The gold antenna substrates were carefully cleaned with acetone, ethanol and MilliQ water followed by a short UV/ozone plasma exposure (between 2-5 min) immediately prior to lipid bilayer deposition. The latter treatment removes any residual organic layer from the gold substrate, guaranteeing a chemically inert and hydrophilic surface.[40] All the steps were carried out under a fume hood. The different lipid mixtures were allowed to dry for roughly an hour in the presence of a weak nitrogen flow, and then kept in vacuum for an additional hour. Consequently, the samples were hydrated in PBS (pH 7.4) and carefully rinsed to remove excess lipids. Samples were imaged and probed by FCS immediately after preparation. All measurements were performed at room temperature.

*Estimation of the illumination areas for the different antenna gaps*

To estimate the illumination areas from our antenna gaps, we considered for the *x*-direction the mean values of the three gap sizes as directly measured from TEM images (Supporting information Figure S3), while for the *y*-direction, we took the distances corresponding to the full-width-at-half-maximum (FWHM) of the respective antenna excitation intensity profiles, as obtained from FDTD simulations (Supporting information Figure S2b-d). The calculated gap areas are $(200 \pm 50)$ nm$^2$, $(1080 \pm 80)$ nm$^2$ and $(2025 \pm 110)$ nm$^2$ for the nominal 10nm, 30nm and 45nm gap sizes, respectively. The sizes of the illumination areas were further calibrated by measuring the diffusion times of the Alexa Fluor 647 dye in solution for five different antenna sizes (10, 25, 30, 35 and 45 nm) considering the reported diffusion coefficient of the dye $(300 \mu m^2/s)$.[45] Results of the calibration are shown in the Supporting information Figure S4. Excellent agreement between the calculated and calibrated values were obtained, except for the smallest antenna (200 nm$^2$ from calculations and 300 nm$^2$ derived from the calibration measurements). We used this experimentally derived illumination area for all the data reported in the ms.

*Fluorescence Microscopy and FCS*

Fluorescence imaging and FCS measurements were performed using a commercial MicroTime 200 setup (PicoQuant). The excitation light of a linearly polarized picosecond laser diode (Pico-Quant LDH-D-C-640) operating at 640 nm in continuous wave mode was focused onto the sample through an Olympus UPLSAPO 60×, 1.2 NA water-immersion objective. A half-wave plate was used to control the polarization of the incoming light. The fluorescence signal was collected through the same objective, separated from the laser light by a dichroic mirror, split by a 50/50 beam splitter cube and sent onto two avalanche photodiodes (APDs) (PicoQuant MPD-50CT) after passing through a 30 μm pinhole conjugated to the focus plane. Two long-pass 650 nm filters were placed in front of the detectors in order to reject backscattered laser light and maximize the fluorescence signal collection. We used two APDs and performed cross-correlation between the two channels instead of autocorrelation of one channel, since it reduces artifacts due to the dead time of each detector and after pulses. A three-axis piezoelectric stage and controller (Physik Instrumente, Karlsruhe, Germany) allowed to scan the sample and precisely position the focus on individual nano-antennas.

FCS measurements were performed by illuminating the sample at an excitation power density of ~2 kW/cm$^2$. The commercial software package SymPhoTime 64 (PicoQuant) was used for the overall handling of the experiment, detection of fluorescent counts, computation of the autocorrelation curves $G(\tau)$ and fitting routines for the FCS analysis. The setup was calibrated by measuring the known three-dimensional diffusion coefficient of Alexa Fluor 647 in solution. Fluorescence time traces on individual nano-antennas were recorded for either 30 or 60 seconds, with a temporal resolution of 4 picoseconds. $G(\tau)$ curves were generated over time windows of typically 10 seconds in length.

The calculated correlation curves $G(\tau)$ were fitted using a two-dimensional Brownian diffusion model, assuming a Gaussian beam profile as shown in Eq. (1).

$$G(\tau) = \sum_{i=0}^{n_{diff}-1} \frac{A_i}{1+\left(\frac{\tau}{\tau_i}\right)} \qquad \text{Eq. (1)}$$

where $n_{diff}$ is the number of independently diffusing species ($n_{diff} = 2$ for the two-component fit used here) and $A_i$ the amplitude of the contribution of the $i^{th}$ diffusing species with the corresponding diffusion time $\tau_i$. We used a two-component fitting since excitation of the dimer antenna inside the nano-aperture leads to two distinct diffusion times as explained in Ref. 27. The shortest diffusion time corresponds to direct excitation at the antenna gap while the second component corresponds to the diffusion times of molecules inside the nano-aperture but away from the antenna hotspot region. These molecules contribute weakly to the overall correlation curve since they are only excited by the residual light inside the nano-aperture (see also Supporting information Figure S5 and Table S2). We also attempted to fit the curves using a three-component fitting but in general the amplitude weight of the third component was either very small (below 3%) and/or rendered a fitting error.

**Supporting Information**

Experimental quality assessment of the four different model lipid bilayers by confocal excitation; In-plane antenna arrays and FDTD simulations; Antenna gap size distribution measured from transmission electron microscopy (TEM) images; Calibration measurements on AlexaFluor 647 in solution to assess the illumination areas of the antenna gaps; Fluorescence intensity time traces and ACF curves for DOPC obtained upon parallel and perpendicularly polarized excitation of the antennas; FCS diffusion plots for the two ternary lipid mixtures at 10 and 20 mol% Chol; Diffusion coefficients of DiD for the different lipid model membrane mixtures as obtained from confocal measurements; Fitting values of the ACF curves for DOPC bilayers at different antenna gap areas.

**Corresponding Author**

*Email: maria.garcia-parajo@icfo.eu, jerome.wenger@fresnel.fr

**Additional Information**

The authors declare no competing financial interests.

## Acknowledgements

We thank K. J. E. Borgman and F. Campelo for fruitful discussions. The research leading to these results has received funding from the European Commission's Seventh Framework Programme (FP7-ICT-2011-7) under grant agreements ERC StG 278242 (ExtendFRET), 288263 (NanoVista), Spanish Ministry of Economy and Competitiveness ("Severo Ochoa" Programme for Centres of Excellence in R&D (SEV-2015-0522) and FIS2014-56107-R), Fundació CELLEX (Barcelona) and CERCA Programme/Generalitat de Catalunya. P.M.W is supported by the ICFOstepstone Fellowship, a COFUND Doctoral Programme of the Marie-Sklodowska-Curie-Action of the European Commission. R.R. is supported by the Erasmus Mundus Doctorate Program Europhotonics (Grant 159224-1-2009-1-FR-ERA MUNDUS-EMJD).

# Supporting Information

Contents:   Figure S1: Experimental quality assessment of the four different model lipid bilayers by confocal excitation.

Figure S2: In-plane antenna arrays and FDTD simulations.

Figure S3: Antenna gap size distribution measured from transmission electron microscopy (TEM) images.

Figure S4. FCS diffusion plot for the free dye Alexa Fluor 647 calibrating the antenna gap areas.

Figure S5: Fluorescence intensity time traces and ACF curves for DOPC obtained upon parallel and perpendicularly polarized excitation of the antennas.

Figure S6: FCS diffusion plots for the two ternary lipid mixtures at 10 and 20 mol% Chol.

Table S1: Diffusion coefficients of DiD for the different lipid model membrane mixtures as obtained from confocal measurements.

Table S2: Fitting of the ACF curves on DOPC bilayers for different antenna gap areas.

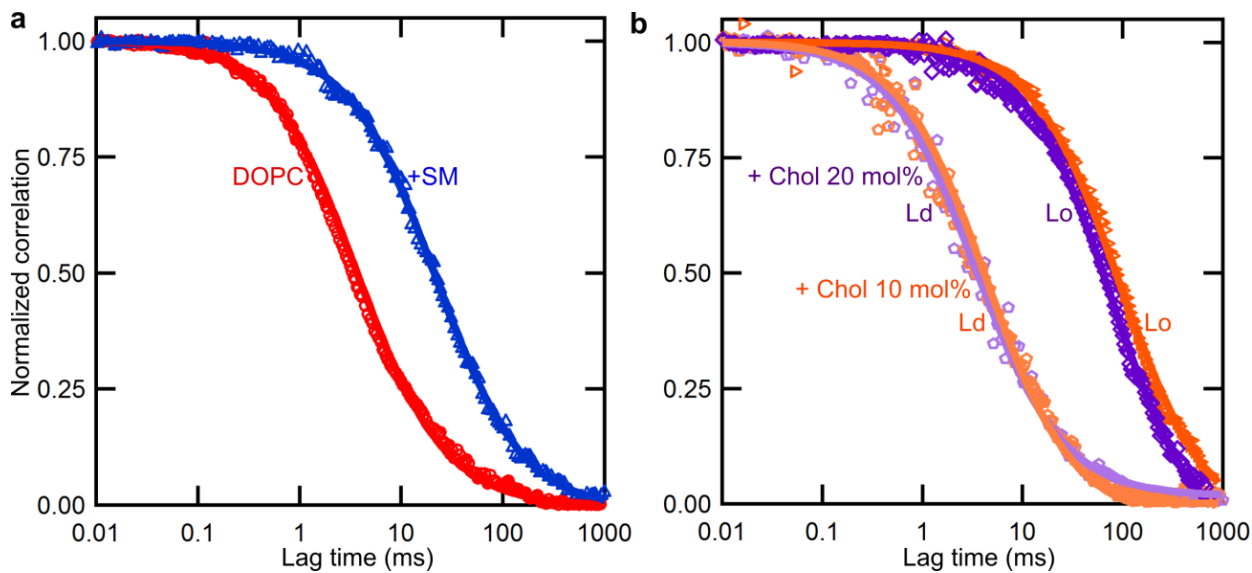

**Figure S1: Experimental quality assessment of the four different model lipid bilayers by confocal excitation.** (a) Normalized autocorrelation functions (ACFs) on pure DOPC and DOPC:SM (1:1) bilayers. (b) Same as (a) but for ternary mixtures of DOPC:SM with the addition of 10 and 20mol% Chol. Upon addition of cholesterol microscopic phase separation into Ld and Lo phases occurs and hence two respective ACFs are shown in (b).

Samples were excited by focusing the laser light ($\lambda$=640nm, ~2 kW/cm$^2$) on to the plane of the membrane using a water-immersion objective (NA=1.2). The fluorescence signal was collected in reflection mode by the same objective, filtered from the excitation light and sent to two single-photon counting APD detectors. Fluorescence fluctuations arising from the diffusion of DiD in the bilayers were recorded for at least 30 seconds at different positions on the sample and the normalized autocorrelation functions (ACF) for each of the four lipid mixtures were calculated.

All the calculated ACFs were fitted using a single component Brownian diffusion model and rendered the characteristic diffusion times $\tau_{diff}$. The respective diffusion coefficients $D$ were then calculated with the approximation valid for 2D diffusion $D \approx \frac{w^2}{4 \times \tau_{diff}}$, where $w$ corresponds to the diffraction limited laser beam waist of $w$~285 nm. Results of the fittings are shown in Table S1. These values compare well to those reported for similar lipid mixtures,[2] validating the quality of the prepared bilayers.

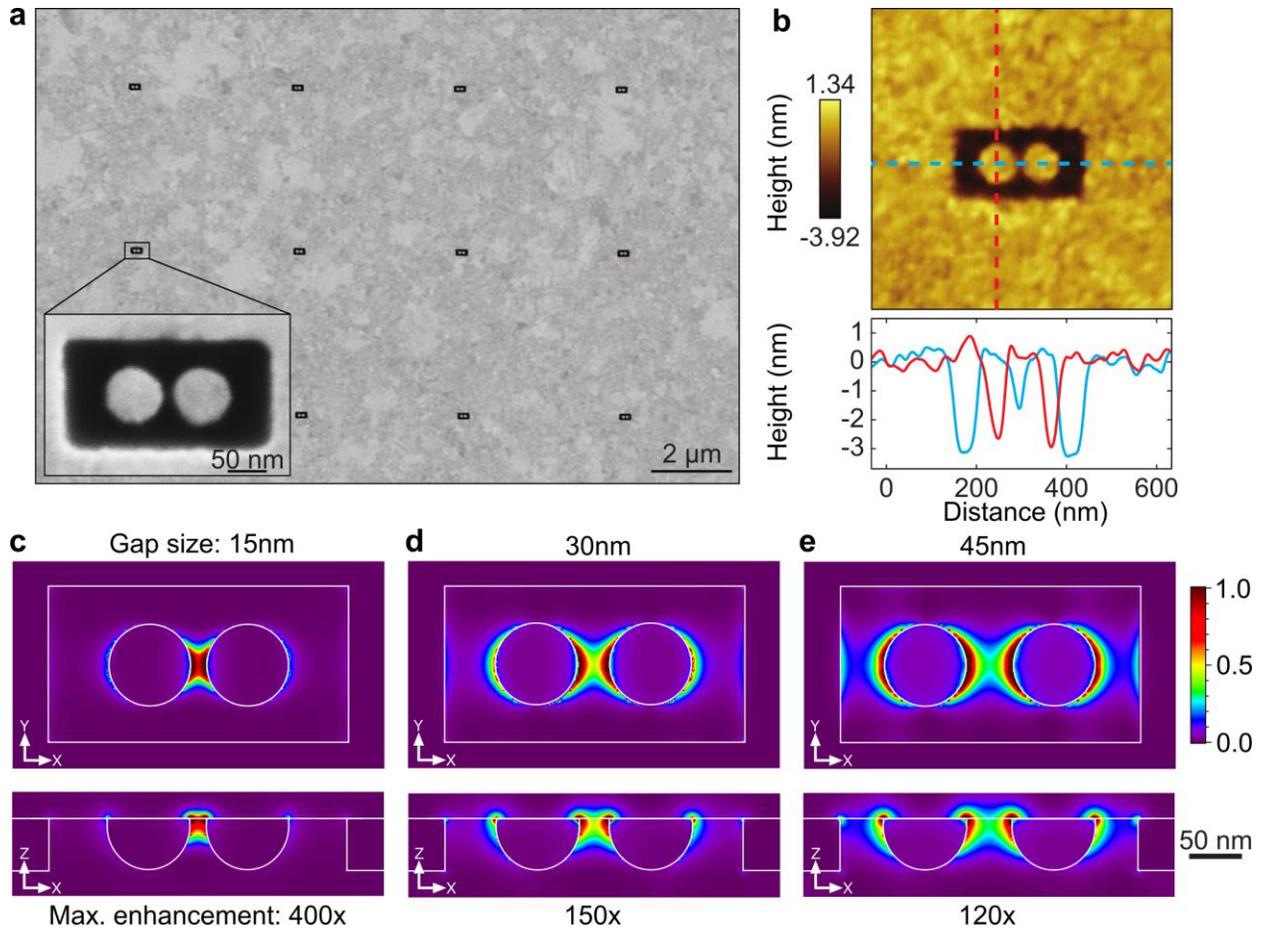

**Figure S2. In-plane antenna arrays and FDTD simulations.** (a) SEM image of part of an array of in-plane dimer antennas embedded in nano-apertures (boxes) with nominal gap size of 10 nm together with an enlarged view of one antenna. The scale bars are 2 μm and 50 nm, respectively. (b) AFM image of one antenna with two line profiles demonstrating the planarity of the antennas. For both profiles (red, blue) the overall AFM amplitude spans 3 nm in height and the horizontal axis is 780 nm. (c-e) The normalized excitation intensity enhancement at 633 nm is computed using the finite-difference time-domain method (FDTD) for a gold nanoantenna (diameter 80 nm) in a nano-aperture for different gap sizes (15, 30 and 45) nm, approximately matching the three antenna gap areas used in the experiments. Computations are performed using the FDTD method (RSoft Fullwave software). The mesh size is 0.5 nm for (c), 1 nm for (d) and 1 nm for (e). We used $2^{14}$ temporal steps of $8.1 \ 10^{-19}$ s. The permittivity of gold is modeled according to the data in Ref. 2. The intensity is recorded at the antenna surface for the cuts in the plane (*y-x*) and along the planes crossing the antenna's center in case of the *z-x* cuts.

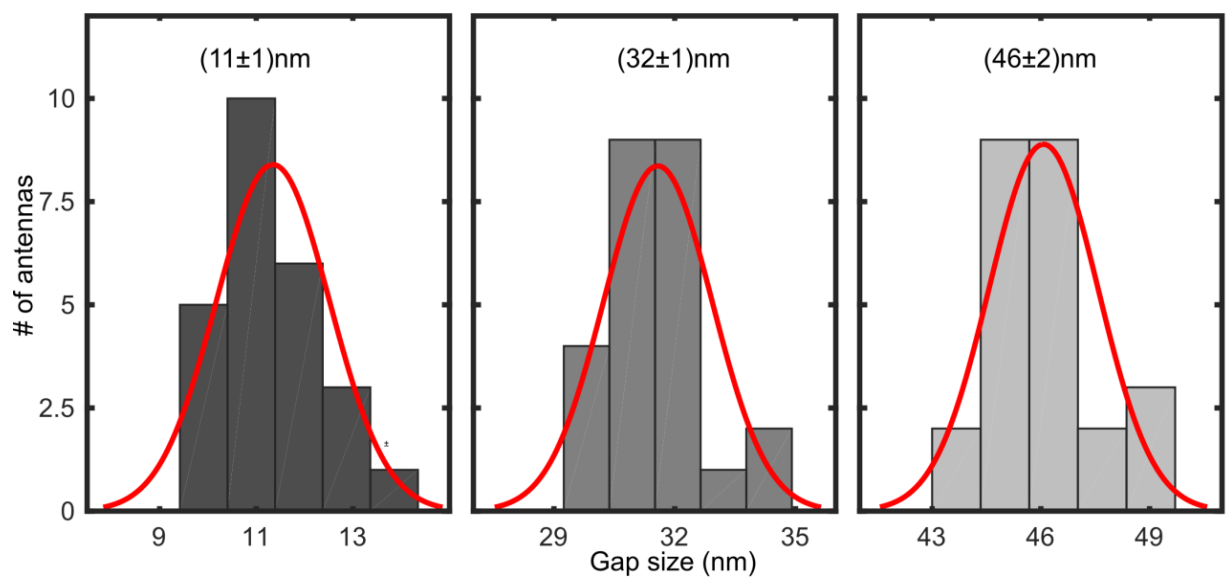

**Figure S3. Antenna gap size distribution measured from transmission electron microscopy (TEM) images.** The plots correspond to distributions of antenna gap sizes of nominally 10, 30 and 45nm sizes (from left to right). Red curves correspond to normal density fittings. The extracted values for the average gap width and associated standard deviation yield (11±1) nm, (32±1) nm and (46±2) nm. Number of antenna gaps measured: 25 for each of the three gaps of nominally 10, 30 and 45nm.

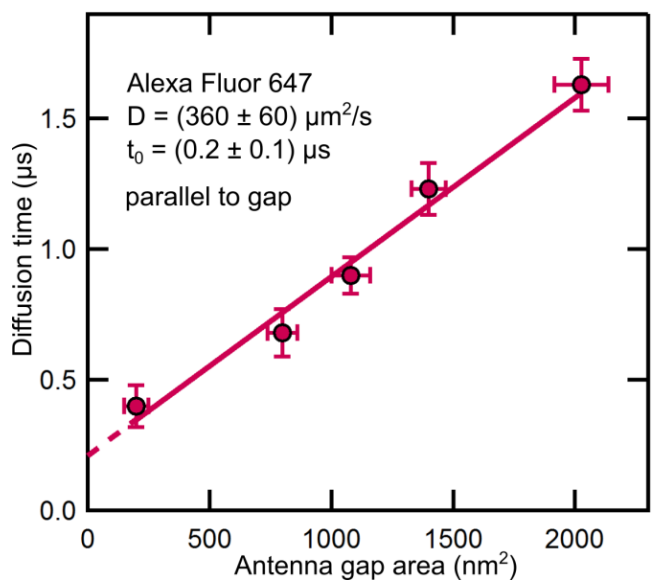

**Figure S4. Calibration curve to obtain the effective illumination areas for the different nano-antenna gap sizes.** Diffusion times versus antenna gap areas resulting from measuring the diffusion of Alexa Fluor 647 dye in solution, considering the reported diffusion coefficient of the dye (300 $\mu m^2/s$) at 21 °C.[3] The excitation intensity at 633 nm wavelength is 2.3 kW/cm$^2$. The probed nano-antennas have nominal gap sizes of 10, 25, 30, 35 and 45 nm. The *calculated* illumination areas resulted in 200, 800, 1080, 1400 and 2025 nm$^2$, respectively. The calibration curve renders similar values for all the antenna gaps, except from the 10 nm antenna. Based on the calibration curve, we then extract the effective illumination area for the 10nm to be of 300 nm$^2$ and used this value further analysis of all the curves reported in this manuscript.

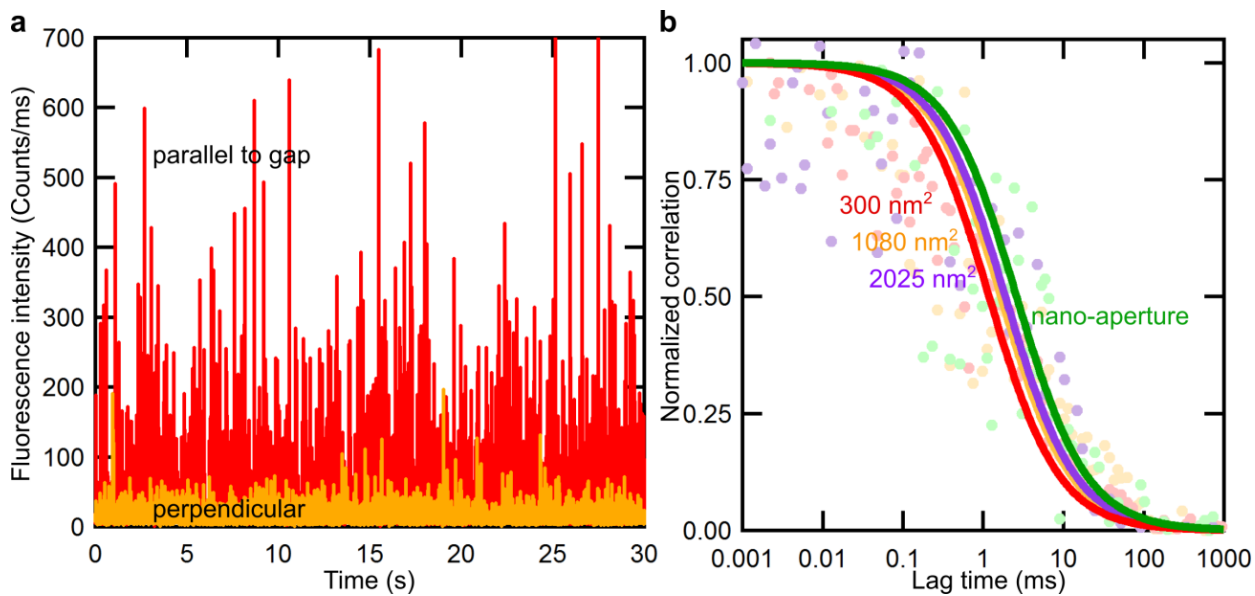

**Figure S5. Fluorescence intensity time traces and ACF curves for DOPC obtained upon parallel and perpendicularly polarized excitation of the antennas.** (a) Representative time traces of DiD in DOPC obtained for the smallest antenna gap area (300 nm$^2$) upon parallel (red) and perpendicular (orange) antenna excitation. A clear reduction in fluorescence intensity is obtained upon perpendicular polarization excitation consistent with the fact that the antenna is out of resonance. The detected intensity corresponds to residual excitation from the nano-aperture alone. (b) Fitting of the ACF curves for perpendicular polarization yields diffusion times of (1.2, 1.65, 1.9) ms for the (300, 1080, 2025) nm$^2$ antenna gap areas respectively, and 2.6 ms for the nano-aperture.

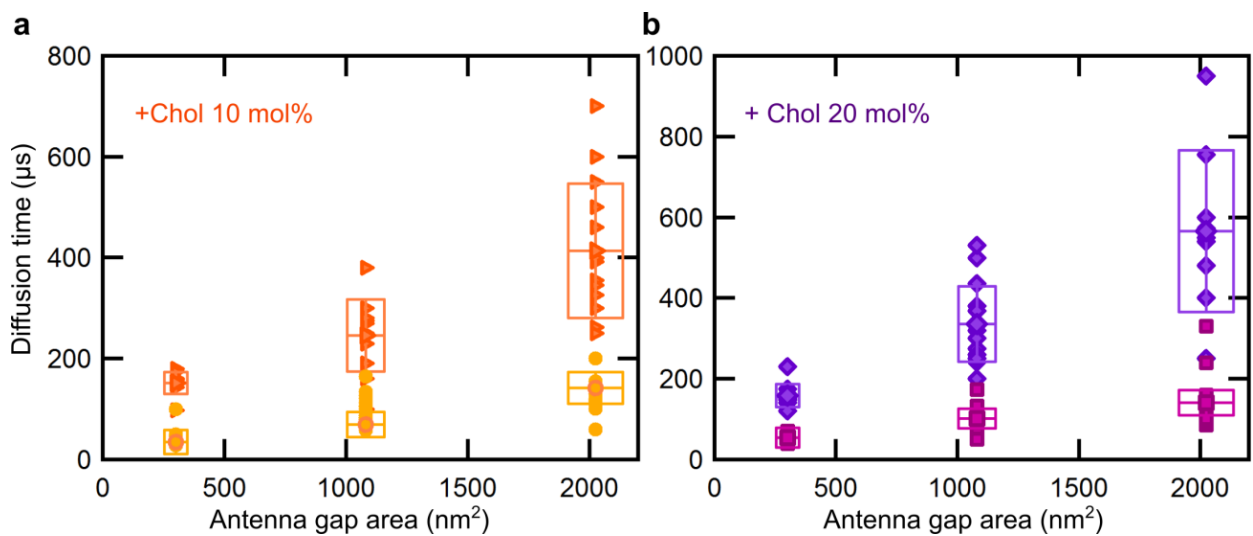

**Figure S6: FCS diffusion plots for the two ternary lipid mixtures at 10 and 20 mol% Chol.** Diffusion times for different gap areas obtained for 10 mol% (a) and 20 mol% Chol (b). The obtained FCS diffusion laws for the two amounts of added Chol show a clear difference in diffusion times for both the Ld (yellow (a) and magenta (b), shorter diffusion times) and the Lo (orange (a) and violet (b), longer diffusion times) phases. Since the diffusion times are clearly different, the assignment of diffusion times to either phase was performed without the need of applying a discriminating threshold.

**Table S1. Diffusion coefficients of DiD for the different lipid model membrane mixtures as obtained from confocal measurements.**

| Lipid mixture | DOPC | DOPC:SM (1:1) | DOPC:SM (1:1) + 10%Chol | | DOPC:SM (1:1) + 20%Chol | |
|---|---|---|---|---|---|---|
| Diffusion coefficient ($\mu m^2$/s) | 5.8±0.3 | 1.7±0.4 | 2.3±0.3 (Ld) | 0.19±0.06 (Lo) | 2.9±0.4 (Ld) | 0.27±0.05 (Lo) |

**Table S2: Fitting of the ACF curves on DOPC bilayers for different antenna gap areas.** Extracted values of the diffusion times $t_{1,2}$ and respective amplitudes $A_{1,2}$ as extracted from the fitting of ACF curves shown in Figure 2b using a two-component Brownian diffusion model. The shortest times ($t_1$) correspond to the diffusion times of DiD through the antenna hotspot regions while $t_2$ corresponds to residual excitation of the dye inside the nano-aperture. The diffusion time values plotted in Figure 2c (and remaining Figures in the main manuscript) correspond to $t_1$.

| | $t_1$ | $A_1$ (%) | $t_2$ | $A_2$ (%) |
|---|---|---|---|---|
| 300 nm$^2$ | (6±1)$\mu$s | 67±3 | (170±15)$\mu$s | 33±8 |
| 1080 nm$^2$ | (25±3)$\mu$s | 79±8 | (1.4±0.2)ms | 21±7 |
| 2025 nm$^2$ | (71±25) $\mu$s | 75±9 | (1.6±0.3)ms | 25±10 |